\documentclass[12pt]{article}
\usepackage{amscd,amssymb,amsmath,latexsym,enumerate}
\usepackage[mathscr]{euscript}
\usepackage{epsfig}
\usepackage{fancybox}
\usepackage{verbatim}

\usepackage{color}

\title{Topological edge states for disordered bosonic systems}

\author{Vittorio Peano$^1$, Hermann Schulz-Baldes$^2$ 
\\
\\
{\small $^1$ Department of Physics, University of Malta, Malta}
\\
{\small $^2$ Department Mathematik, Friedrich-Alexander-Universit\"at Erlangen-N\"urnberg, Germany}
}

\date{ }

\newtheorem{theo}{Theorem}

\newtheorem{proposi}[theo]{Proposition}

\newcommand{\BM}{{\mathbb B}}
\newcommand{\CM}{{\mathbb C}}
\newcommand{\NM}{{\mathbb N}}
\newcommand{\RM}{{\mathbb R}}
\newcommand{\SM}{{\mathbb S}}

\newcommand{\ZM}{{\mathbb Z}}
\newcommand{\PM}{{\mathbb P}}

\newcommand{\HM}{{\mathbb H}}

\newcommand{\UM}{{\mathbb U}}

\newcommand{\Aa}{{\cal A}}

\newcommand{\NN}{{\bf N}}

\newcommand{\HH}{{\bf H}}

\newcommand{\EE}{{\bf E}}

\newcommand{\Ee}{{\cal E}}
\newcommand{\Ff}{{\cal F}}

\newcommand{\Tr}{\mbox{\rm Tr}}
\newcommand{\Tt}{{\cal T}}

\newcommand{\Kk}{{\cal K}}
\newcommand{\Hh}{{\cal H}}

\newcommand{\one}{{\bf 1}}

\newcommand{\TR}{{\rm Tr}} 
\newcommand{\Ch}{{\rm Ch}} 
\newcommand{\diag}{{\rm diag}} 
\newcommand{\trans}{{t}}

\newcommand{\ph}{{\mbox{\rm\tiny ph}}}
\newcommand{\ess}{{\mbox{\rm\tiny ess}}}

\newcommand{\anni}{{\mathfrak a}}
\newcommand{\crea}{{\mathfrak a}^*}

\begin{document}

\maketitle

\begin{abstract}
Quadratic bosonic Hamiltonians over a one-particle Hilbert space can be described by a Bogoliubov-de Gennes (BdG) Hamiltonian on a particle-hole Hilbert space.  In general, the BdG Hamiltonian is not selfadjoint, but only $J$-selfadjoint on the particle-hole space viewed as a Krein space. Nevertheless, its energy bands can have non-trivial topological invariants like Chern numbers or winding numbers. By a thorough analysis for tight-binding models, it is proved that these invariants lead to bosonic edge modes which are robust to a large class of possibly disordered perturbations. Furthermore, general scenarios are presented for these edge states to be dynamically unstable, even though the bulk modes are stable.


%
\end{abstract}

\section{Introduction}
\label{sec-overview}

Robust topological edge modes are a well-established feature of topological electron systems and there is a continued theoretical and experimental effort to further explore and use these states. During the last decade it has been realized that the phenomena of topologically protected states also appears in other physical systems described by wave equations. Examples range from cold atom systems \cite{JMD,ALS,GBZ}, to photonic \cite{RH,LSJ,DL}, phononic \cite{PP,KL}, and magnonic systems \cite{KNL,SMMO}. In this mathematical contribution, we consider quadratic bosonic Hamiltonians over a tight-binding Hilbert space and analyze their topological invariants and associated edge states. Our analysis goes beyond a strictly single-particle framework by including bosonic pairing terms.  Such terms describe the simultaneous creation or annihilation of pairs of entangled quasiparticles and play an important role in different  topological settings including  parametrically driven systems \cite{SKC,PHMC,PHB}, the mean-field analysis of Bose-Einstein condensates \cite{Bar,GLB,EB,BKRL,EBPB,WZK}, and magnonic crystals \cite{SMMO}. Here is what is achieved in this paper:

\begin{itemize}

\item A thorough, albeit basic discussion of quadratic bosonic Hamiltonians and associated Bogoliubov-de Gennes (BdG) Hamiltonians generating the dynamics. The BdG Hamiltonian is {\it not} selfadjoint, but rather a Real $J$-selfadjoint operator on a Krein space with Real structure, which is in fact just the standard particle-hole Hilbert space. The role of this Krein space structure for the spectral analysis of the BdG Hamiltonian is stressed and explored (Section~\ref{sec-BdGgen}).

\item A general definition of topological invariants for  covariant tight-binding BdG Hamiltonians is given and basic properties of the invariants are derived (Section~\ref{sec-TopInvBdG}). These topological invariants represent extensions of those defined in Ref.~\cite{SMMO}, both to disordered systems as well as to amorphous systems as recently studied in \cite{AS,MNHTI}.
 
\item The invariants have no direct physical meaning, basically because the equivalent of the Fermi projection (an indicator function of the BdG Hamiltonian which is an idempotent, but not selfadjoint) is of little interest in a bosonic system. However, by the bulk-boundary correspondence, non-trivial invariants imply the existence of topological edge states. This is proved in detail for two-dimensional systems which are thermodynamically stable, by applying mathematical tools from fermionic topological insulators (Section~\ref{sec-EdgeThermoStab}).

\item For dynamically stable bulk Hamiltonians (namely, bulk BdG Hamiltonians with real spectrum, see Section~\ref{sec-DynStab}), it is shown how to produce unstable edge or bound states (more precisely: these BdG Hamiltonians over a half-space have spectrum off the real axis). This conceptualizes earlier works \cite{Bar,GLB,PHMC} (Section~\ref{sec-UnstableBound}).

\end{itemize}




\noindent {\bf Acknowledgements:} Part of the results of this paper were obtained in collaboration with F.~Marquardt and S. Kaul during the preparation of the master thesis of S.~Kaul. This work is partially supported by the DFG. VP acknowledges support by the Julian Schwinger Foundation and the EU HOT network.

\section{Generalities on bosonic BdG Hamiltonians}
\label{sec-BdGgen}

This section recollects a few basic and well-known general facts on quadratic Hamiltonians on bosonic Fock space, many of which can be found in standard text books like \cite{BR}.  Apart from fixing notations and terminology, this section hopefully allows the reader to place the results below in the more general framework of interacting systems. Any technical difficulties linked to the unboundedness of the bosonic creation and annihilation operators are intentionally swept under the carpet (an excellent standard reference is \cite{BrR}). On the other hand, facts about the various operators on the particle-hole Hilbert space are rigorous. In fact, starting from Section~\ref{sec-TopInvBdG} below, we will only work with bounded linear operators on this particle-hole Hilbert space and hence essentially in a one-particle framework. None of the statements made there require the notion of Fock space or operators thereon.

\subsection{Quadratic bosonic Hamiltonian}
\label{sec-QuadraticHam}

Let us consider the separable complex one-particle Hilbert space $\Hh\cong\ell^2(\{1,\ldots,N\})$ with a designated orthonormal basis  $|n\rangle$ with $n\geq 1$. It will be equipped with a real structure denoted simply by a complex conjugation bar. We will be interested in both the case of finite and infinite $N$. Starting from Section~\ref{sec-TopInvBdG} below, $\Hh$ will be the tight-binding Hilbert space, but this spatial structure is not exploited in this section.  Associated to $\Hh$ is the bosonic Fock space $\Ff=\Ff(\Hh)$. The creation and annihilation operators in the state $|n\rangle$ are denoted by $\crea_{n}$ and $\anni_{n}$ respectively. They satisfy the canonical commutation relation (CCR):
$$
[\anni_{n},\anni_{n'}]\;=\;0\;,
\qquad
[\crea_{n},\crea_{n'}]\;=\;0\;,
\qquad
[\anni_{n},\crea_{n'}]\;=\;\delta_{n,n'}
\;.
$$
Using Weyl operators, one can construct the CCR C$^*$-algebra $\mathfrak{A}$, and then the unbounded operators $\anni_n$ and $\crea_n$ are affiliated to $\mathfrak{A}$ \cite{BrR}.  It will be useful to introduce column vectors of annihilation and creation vectors:
$$
\anni
\;=\;
\begin{pmatrix}
\anni_1 \\ \anni_2 \\ \vdots
\end{pmatrix}
\;,
\qquad
\crea
\;=\;
\begin{pmatrix}
\crea_1 \\ \crea_2 \\ \vdots
\end{pmatrix}
\;.
$$ 
Their transposes $\anni^\trans $ and $(\crea)^\trans $ are line vectors. If then $\phi$ and $\psi$ are column vectors of complex numbers, we use the notations
$$
\psi^\trans \anni
\;=\;
\sum_{n\geq 1} \psi_n\anni_n
\;,
\qquad
\phi^\trans \crea
\;=\;
\sum_{n\geq 1} \phi_n\crea_n
\;,
\qquad
\begin{pmatrix} \psi \\ \phi \end{pmatrix}^\trans \begin{pmatrix} \anni \\ \crea \end{pmatrix}
\;=\;
\psi^\trans \anni
\,+\,
\phi^\trans \crea
\;.
$$
Of course, if there are infinitely many entries, some care is needed at this point, but we neglect these issues here. The above notations allow to rewrite the CCR's in a compact form:
\begin{equation}
\label{eq-CCR}
\left[
\begin{pmatrix} \psi \\ \phi \end{pmatrix}^\trans \begin{pmatrix} \anni \\ \crea \end{pmatrix}
,
\begin{pmatrix} \psi' \\ \phi' \end{pmatrix}^\trans \begin{pmatrix} \anni \\ \crea \end{pmatrix}
\right]
\;=\;
\begin{pmatrix} \psi \\ \phi \end{pmatrix}^\trans 
\begin{pmatrix}
0 & \one \\ -\one & 0
\end{pmatrix}
\begin{pmatrix} \psi' \\ \phi' \end{pmatrix}
\;.
\end{equation}
Also the number operator on Fock space takes the simple form 
\begin{equation}
\label{eq-NumberOp}
\NN
\;=\;
\frac{1}{2}\;(\crea)^\trans \,\anni
\;+\;
\frac{1}{2}\;\anni^\trans \,\crea
\;,
\end{equation}
Using the CCR's this can be rewritten as $\NN=(\crea)^\trans \,\anni+\frac{1}{2} N$, but the constant term $\frac{1}{2}N$ may, of course, be infinite. In this work we consider a generic formally selfadjoint Hamiltonian ${\bf H}$ on $\Ff$ which is quadratic in the bosonic creation and annihilation operators: 
\begin{equation}
\label{eq-HBdG}
{\bf H}
\;=\;
\frac{1}{2}\;(\crea)^\trans \,h\,\anni
\;+\;
\frac{1}{2}\;\anni^\trans \,\overline{h}\,\crea
\;+\;
\frac{1}{2}\;(\crea)^\trans \,\Delta\,\crea
\;+\;
\frac{1}{2}\;\anni^\trans \,\Delta^*\;\anni
\;.
\end{equation}
Here $h=h^*=(h(n,n'))_{n,n'\geq 1} $ and $\Delta=(\Delta(n,n'))_{n,n'\geq 1} $ are bounded linear operators on $\Hh$, and the following notation has been used 
\begin{equation}
\label{eq-HBdGsum}
(\crea)^\trans \,h\,\anni
\;=\;
\sum_{n,n'\geq 1}\,\crea_{n}\,h(n,n')\, \anni_{n'}
\;,
\end{equation}
and similarly for the three other terms in \eqref{eq-HBdG}.  Furthermore, $h^*$, $\overline{h}$ and $h^\trans $ denote the adjoint, complex conjugate and transpose of $h$, respectively. Let us note that the CCR's imply
$$
(\crea)^\trans \,h\,\anni
\;=\;
\anni^\trans \,h^\trans \,\crea\,-\,\Tr(h)\,\one
\;,
\qquad
(\crea)^\trans \,\Delta\,\crea\;=\;(\crea)^\trans \,\Delta^\trans \,\crea
\;.
$$
Hence up to the (again possibly infinite) constant term $\Tr(h)\,\one$, the two first summands in \eqref{eq-HBdG} can be regrouped to $(\crea)^\trans h\anni$. Moreover, it is no restriction to assume 
$$
\Delta\;=\;\Delta^\trans 
\;,
$$ 
because any antisymmetric component of $\Delta$ would give no contribution to ${\bf H}$.  Using a block matrix representation, the Hamiltonian ${\bf H}$ in \eqref{eq-HBdG} can be written in the form
\begin{equation}
\label{eq-HBdG2}
{\bf H}
\;=\;
\frac{1}{2}\;
\begin{pmatrix} \crea \\ \anni \end{pmatrix}^\trans 
A
\begin{pmatrix} \anni \\ \crea \end{pmatrix}
\;,
\qquad
A
\;=\;
\begin{pmatrix}
h & \Delta \\ \overline{\Delta} & \overline{h}
\end{pmatrix}
\;=\;
\begin{pmatrix}
h & \Delta \\ \Delta^* & \overline{h}
\end{pmatrix}
\;.
\end{equation}
The linear operator $A=A^*$ acts on the so-called particle-hole Hilbert space $\Hh_{\mbox{\tiny\rm ph}}=\Hh\otimes\CM^2_{\mbox{\tiny\rm ph}}$ and in a large part of the mathematics and physics literature \cite{BR,SMMO,Bar,BB,NNS} it is called the Bogoliubov-de Gennes (BdG) Hamiltonian, but here this name is reserved for the generator of the dynamics introduced in Section~\ref{sec-PHSBdG}. As $h$ and $\Delta$ are bounded operators, clearly also $A$ is bounded. This does not automatically mean that $\HH$ is also a well-defined operator on Fock space. One instance where this is straightforward is described in the next section.

\subsection{Thermodynamical stability}
\label{sec-Thermodyn}

If $A\geq 0$ is positive semidefinite, then also $\HH$ is positive semidefinite and can be defined as a selfadjoint operator on Fock space via the Friedrichs extension, see \cite{BB,NNS} and references therein. On the other hand, if $A$ is neither positive nor negative semidefinite, more care is needed when properly defining the operator $\HH$. From a physical point of view, negative spectrum of $A$ leads to states that can be populated with more and more particles while the energy is lowered further and further (for fermions this is not possible). In particular, this leads to a {\em thermodynamically unstable} system, and, by contradistinction, a Hamiltonian $\HH$ is called {\em thermodynamically stable} if $A\geq 0$. Hence the hypothesis $A\geq 0$ seems reasonable and possibly even necessary on physical grounds, but on the other hand, one can also be interested in driven systems for which thermodynamical stability does not hold. Nevertheless, one can use $\HH$ as a generator of a dynamics on the CCR algebra which can, moreover, be dynamically stable in the sense described in Section~\ref{sec-TimeEvolve} below.

\vspace{.2cm}

A simple way to assure thermodynamical stability is to add a chemical potential $\mu\in\RM$, namely replace the Hamiltonian by
\begin{equation}
\label{eq-BdGmu}
\HH_\mu
\;=\;
\HH\,-\,\mu\,\NN
\;=\;
\frac{1}{2}\;
\begin{pmatrix} \crea \\ \anni \end{pmatrix}^\trans 
A_\mu
\begin{pmatrix} \anni \\ \crea \end{pmatrix}
\;,
\qquad
A_\mu\;=\;A\,-\,\mu\,\one
\;.
\end{equation}
Choosing $\mu$ sufficiently negative, one has $A_\mu\geq 0$. Also sufficiently large values of $\mu$ lead to a negative definite $A_\mu$, whenever $A$ is bounded. In the following, $\mu$ will be seen as a parameter that can be changed in an experimental set-up.

\subsection{Particle-hole symmetric BdG Hamiltonian}
\label{sec-PHSBdG}

While the representation \eqref{eq-HBdG2} using a selfadjoint operator $A$ on $\Hh_\ph$ is appealing and widely used, there are also good reasons to rather proceed otherwise and this is described next. Let us introduce four auxiliary operators on $\Hh_\ph=\Hh\otimes\CM^2_\ph$:
$$
J
\;=\;
\begin{pmatrix} \one & 0 \\ 0 & -\one
\end{pmatrix}\;,
\qquad
K
\;=\;
\begin{pmatrix} 0 & \one \\ \one & 0
\end{pmatrix}\;,
\qquad
I
\;=\;
\begin{pmatrix} 0 & - \one \\ \one & 0
\end{pmatrix}
\;,
\qquad
C\;=\;
\frac{1}{\sqrt{2}}
\begin{pmatrix}
\one & - \imath \\ \one & \imath
\end{pmatrix}
\;.
$$
They are all unitary. The first three are real and square to the identity or its negative, and are essentially the anti-commuting Pauli matrices. The last one is the Cayley transform. These operators will be used to implement symmetries. For example, the operator $A$ defined in \eqref{eq-HBdG2} clearly satisfies
\begin{equation}
\label{eq-BdGsymmetry0}
K\,\overline{A}\,K
\;=\;
A
\;.
\end{equation}
Of course, for functions $f(A)$ of the selfadjoint operator $A$, a corresponding property $K\overline{f(A)}K=f(A)$ holds. Now let us come to the main object of this subsection. Instead of \eqref{eq-HBdG2}, one can clearly also write
\begin{equation}
\label{eq-HBdG3}
{\bf H}
\;=\;
\frac{1}{2}\;
\begin{pmatrix} \crea \\ -\anni \end{pmatrix}^\trans 
H
\begin{pmatrix} \anni \\ \crea \end{pmatrix}
\;,
\qquad
H
\;=\;
J\,A
\;=\;
\begin{pmatrix}
h & \Delta \\ -\overline{\Delta} & -\overline{h}
\end{pmatrix}
\;.
\end{equation}
The operator $H$ then has a particle-hole symmetry (PHS), in complete analogy to the fermionic BdG Hamiltonian \cite{BR,Zir}:
\begin{equation}
\label{eq-BdGsymmetry}
K^*\,\overline{H}\,K
\;=\;
-\,H
\;.
\end{equation}
This is algebraically equivalent to \eqref{eq-BdGsymmetry0} because $JK=-KJ$. For this reason, we refer to $H$ as the BdG Hamiltonian, as in \cite{Zir}. An apparent disadvantage is that $H$ is {\it not} selfadjoint, unless $\Delta$ is purely imaginary. The BdG Hamiltonian is, however, a bounded $J$-selfadjoint operator on the Krein space $(\Hh_\ph,J)$, namely 
\begin{equation}
\label{eq-JHermition}
H^*
\;=\;J\,H\,J
\;,
\end{equation}
on top of having the PHS \eqref{eq-BdGsymmetry}. Section~\ref{sec-TimeEvolve} explains the main reason why $H$ and its spectral analysis is relevant. Let us briefly comment on what happens when a chemical potential $\mu\not=0$ is present as in \eqref{eq-BdGmu}. Then the BdG Hamiltonian becomes
\begin{equation}
\label{eq-JHermitianMu}
H_\mu\;=\;J\,A_\mu
\;=\;
H\,-\,\mu\,J
\;.
\end{equation}
It is still a Real $J$-selfadjoint, namely $H_\mu$ satisfies \eqref{eq-BdGsymmetry} and \eqref{eq-JHermition}.

\subsection{Krein space perspective}
\label{sec-KreinPer}

In this section, we collect a few facts related to the symmetries of $H$ (and also $H_\mu$) by referring to general properties of operators on Krein spaces with Real symmetries \cite{SV}. A Krein space $(\Kk,J)$ is a Hilbert space $\Kk$ equipped with a so-called fundamental symmetry $J$, a linear operator satisfying $J^2=\one$ and $J=J^*$. It is no restriction to assume that the Krein space is of the form $(\Hh_\ph,J)$ with $J$ as described above. It becomes a Real Krein space if there is a real structure (denoted by a complex conjugation bar) and a real unitary $K$ which either squares to $\one$ or $-\one$ and either commutes or anti-commutes with $J$. Here there are four kinds of Real Krein spaces. Here $K^2=\one$ and $KJ=-JK$ so that we deal with a Real Krein space of kind $(1,-1)$ in the terminology \cite{SV}.  In this context, the identity \eqref{eq-BdGsymmetry} can be interpreted as a Real symmetry of $J$-selfadjoint $H$ on the Real Krein space $(\Hh_\ph,J,K)$. Otherwise stated, $H$ is an element of the infinite dimensional Lie algebra
$$
\HM(\Hh_\ph,J,K)
\;=\;
\left\{
H\in\BM(\Hh_\ph)
\;:\;
H^*=JHJ\;,\;\;K\overline{H}K=-H
\right\}
\;.
$$
Inversely, one readily checks
$$
\HM(\Hh_\ph,J,K)
\;=\;
\left\{
\begin{pmatrix}
h & \Delta \\ -\overline{\Delta} & -\overline{h}
\end{pmatrix}
\;:\;
h=h^*\in\BM(\Hh)\;,\;\;\Delta=\Delta^\trans \in\BM(\Hh)
\right\}
\;.
$$
The associated Lie group, a subgroup of the general linear group $\mbox{\rm GL}(\Hh_\ph)$ on $\Hh_\ph$, is obtained upon exponentiation $T=e^{\imath H}$:
$$
\UM(\Hh_\ph,J,K)
\;=\;
\left\{
T\in\mbox{\rm GL}(\Hh_\ph)
\;:\;
T^*JT=J\;,\;\;K\overline{T}K=T
\right\}
\;.
$$
Operators therein are said to be $J$-unitary with Real symmetry $K\overline{T}K=T$.  Note that one then also has $T^\trans IT=I$ and that the relation $TJT^*=J$ follows upon inversion of $T^*JT=J$ so that $\UM(\Hh_\ph,J,K)$ is invariant under the adjoint operation. Just as its Lie algebra, this group can be written out more explicitly:
$$
\UM(\Hh_\ph,J,K)
\;=\;
\left\{
\begin{pmatrix}
u & v \\ \overline{v} & \overline{u}
\end{pmatrix}
\;:\;
u,v\in\mbox{\rm GL}(\Hh)\;,\;\;u^*u-v^\trans \overline{v}=\one\;,\;\;u^*v=v^\trans \overline{u}
\right\}
\;.
$$
After a Cayley transform, this group takes a different form:
$$
C^*\,\UM(\Hh_\ph,J,K)\,C
\;=\;
\left\{
T\in\mbox{\rm GL}(\Hh_\ph)
\;:\;
T^*IT=I\;,\;\;\overline{T}=T
\right\}
\;.
$$
In the finite dimensional case $N=\dim(\Hh)<\infty$, this latter is the real symplectic group SP$(2N,\RM)$.  As $H\in \HM(\Hh_\ph,J,K)$, all techniques of the spectral analysis for $J$-selfadjoint operators with Real symmetry \cite{SV} apply to the BdG Hamiltonian. For example, the PHS \eqref{eq-BdGsymmetry} and the $J$-selfadjointness \eqref{eq-JHermition} imply respectively  that the spectrum of $H$ satisfies
\begin{equation}
\label{eq-FourFold}
\sigma(H)\;=\;-\,\overline{\sigma(H)}\;=\;\overline{\sigma(H)}\;=\;-\,{\sigma(H)}
\;.
\end{equation}
The generic scenarios for eigenvalues to leave the real axis are given by a quadruple Krein collision (via two doubly degenerate eigenvalue pairs with opposite sign) or a tangent bifurcation (via a doubly degenerate eigenvalue $0$) \cite{SV}, see the discussion in Section~\ref{sec-DynStab}.

\vspace{.2cm}

Finally let us collect some properties for functions of $H\in\HM(\Hh_\ph,J,K)$. First of all, \eqref{eq-JHermition} and \eqref{eq-BdGsymmetry} imply respectively that for any real-valued polynomial $p:\RM\to\RM$, one has $p(H)^*=Jp(H)J$, and for even and odd real polynomials $p_\pm$ one has $K^*p_\pm (\overline{H})K=\pm\,p_\pm(H)$. If functional calculus for other functions is given (see Section~\ref{sec-Bogo}), then these symmetry properties naturally transpose. Furthermore, let $\Gamma$ be a closed curve in the resolvent set of $H$. Then there is an associated Riesz projection $Q_\Gamma$ given by holomorphic functional calculus: 
$$
Q_\Gamma
\;=\;
\oint_\Gamma\frac{dz}{2\pi\imath}\;(z\,\one-H)^{-1}
\;.
$$
On top of the usual properties of Riesz projections, one readily verifies as in \cite{SV} that also 
\begin{equation}
\label{eq-Q*}
Q_\Gamma^*
\;=\;
J\,Q_\Gamma\,J
\;,
\qquad
K\,\overline{Q_\Gamma}\,K
\;=\;
Q_{-\overline{\Gamma}}
\;,
\end{equation}
where $-\overline{\Gamma}$ is the inverted and reflected curve (obtained by taking the negative and complex conjugate in the complex plane). This curve is also positively oriented and lies in the resolvent set due to \eqref{eq-FourFold}. Let us stress that these projections are, in general, not selfadjoint, namely they are merely idempotents. One instance in which they are orthogonal projections is when the Hamiltonian is diagonal, that is $\Delta=0$.

\subsection{Time evolution}
\label{sec-TimeEvolve}

In this section, it will be shown that the dynamics generated by a bosonic quadratic Hamiltonian is determined by its BdG Hamiltonian $H$, and {\em not} the selfadjoint operator $A$. In general, the time evolution of an observable ${\bf B}$ on Fock space $\Ff$ is determined by Hamilton's equation
\begin{equation}
\label{eq-Heisen}
\partial_t {\bf B}
\;=\;
\imath [\HH,{\bf B}]
\;.
\end{equation}
If now $\HH$ is quadratic as in \eqref{eq-HBdG} or \eqref{eq-HBdG2}, then this dynamics leaves the span of operators built out of products of $M$ creation and annihilation operators invariant,  for each $M\geq 1$. Indeed, due to the CCR's, the commutator on the r.h.s. of \eqref{eq-Heisen} then lies again in this span. To determine this dynamics, let us write such as ${\bf B}$ in the form
\begin{equation}
\label{eq-TensorRep}
{\bf B}
\;=\;
B\;
\begin{pmatrix} \anni \\ \crea \end{pmatrix}
\cdots
\begin{pmatrix} \anni \\ \crea \end{pmatrix}
\;=\;
\sum_{n_1,\ldots,n_M=1}^{2N}
B^{n_1,\ldots,n_M}\;
\begin{pmatrix} \anni \\ \crea \end{pmatrix}_{n_1}
\cdots
\begin{pmatrix} \anni \\ \crea \end{pmatrix}_{n_M}
\;,
\end{equation}
where $B$ is a tensor of degree $(0,M)$ over $\Hh_\ph$. If $\dim(\Hh)=N=\infty$, then some summability condition on the tensor has to be imposed, but, as already stated in the introduction to this chapter, we are deliberately vague about such issues here. Now by \eqref{eq-Heisen},
$$
\partial_t{\bf B}
\;=\;
\sum_{m=1}^M
B\;
\begin{pmatrix} \anni \\ \crea \end{pmatrix}
\cdots
\begin{pmatrix} \anni \\ \crea \end{pmatrix}
\,
\imath\Big[\HH,\begin{pmatrix} \anni \\ \crea \end{pmatrix}\Big]
\begin{pmatrix} \anni \\ \crea \end{pmatrix}
\cdots
\begin{pmatrix} \anni \\ \crea \end{pmatrix}
\;.
$$
Hence it is sufficient to calculate the commutator with operators that are linear in $\anni$ and $\crea$. Replacing \eqref{eq-HBdG} and using the CCR's, careful tracking of signs leads to the crucial identity
\begin{equation}
\label{eq-LinEvolve}
\partial_t
\begin{pmatrix} \anni \\ \crea \end{pmatrix}
\;=\;
\imath\,\Big[\HH,\begin{pmatrix} \anni \\ \crea \end{pmatrix}\Big]
\;=\;
-\,\imath\,H\,\begin{pmatrix} \anni \\ \crea \end{pmatrix}
\;.
\end{equation}
Thus, for ${\bf B}$ as above,
\begin{equation}
\label{eq-TensorEvolve}
\partial_t{\bf B}
\;=\;
-\,\imath
\sum_{m=1}^M
B\;r_m(H)
\;
\begin{pmatrix} \anni \\ \crea \end{pmatrix}
\cdots
\begin{pmatrix} \anni \\ \crea \end{pmatrix}
\;,
\end{equation}
where $r_m(H)$ is the right multiplication on the $m$th entry of $B$ with $H$ viewed as tensor of degree $(1,1)$. Equation \eqref{eq-TensorEvolve} can also be read as an evolution equation for the components of the tensor $B$. In conclusion, this establishes the claim made in the first phrase of this subsection. 

\vspace{.2cm}

Of particular interest are the observables which are quadratic in $\anni$ and $\crea$, that is, the case $M=2$. Instead of using the representation \eqref{eq-TensorRep}, it is convenient to rather write such an observable exactly in same way as the BdG Hamiltonian in \eqref{eq-HBdG3}:
\begin{equation}
\label{eq-TensorM=2}
{\bf B}
\;=\;
\frac{1}{2}\;
\begin{pmatrix} \crea \\ -\anni \end{pmatrix}^\trans 
\widetilde{B}
\begin{pmatrix} \anni \\ \crea \end{pmatrix}
\;.
\end{equation}
Of course, the components of the $(1,1)$-tensor $\widetilde{B}$ can readily be expressed in terms of those of the $(0,2)$-tensor $B$. Under this identification, \eqref{eq-TensorEvolve} becomes 
$$
\partial_t \widetilde{B}
\;=\;
\imath[H,\widetilde{B}]
\;.
$$
This can also be verified by another direct calculation from scratch. Hence for quadratic observables, the dynamics under a quadratic Hamiltonian $\HH$ given by Heisenberg equation \eqref{eq-Heisen} is again given by the Heisenberg equation on the particle-hole Hilbert space associated to the BdG Hamiltonian $H$, provided that one uses the representation \eqref{eq-TensorM=2} for the quadratic observable. The algebraic reason for this is the following fact which has a fermionic analog, {\it e.g.} \cite{Zir,DS3}, and can be verified in a similar manner as the above Heisenberg equation on particle-hole space.

\begin{proposi}
\label{prop-BosonCommutator}
The map $\widetilde{B}\mapsto {\bf B}$ defined by \eqref{eq-TensorM=2} is a Lie-algebra homomorphism, namely
\begin{equation}
\label{eq-commutatorid}
[{\bf B},{\bf B'}]
\,=\,
\frac{1}{2}\,
\begin{pmatrix} \crea \\ - \anni\end{pmatrix}^\trans 
[\widetilde{B},\widetilde{B'}]
\begin{pmatrix} \anni \\ \crea\end{pmatrix}
\;.
\end{equation}
\end{proposi}

\subsection{Dynamical stability and instability}
\label{sec-DynStab}

The linear time-evolution can be solved as usual by exponentiation. For example, the solution of \eqref{eq-LinEvolve} is given by
$$
\begin{pmatrix} \anni \\ \crea \end{pmatrix}(t)
\;=\;
e^{-\imath tH}\,\begin{pmatrix} \anni \\ \crea \end{pmatrix}
\;.
$$
The exponential $e^{-\imath t H}$ lies in the Lie group $\UM(\Hh_\ph,J,K)$ of the Lie algebra $\HM(\Hh_\ph,J,K)$, see Section~\ref{sec-KreinPer}. Clearly, the spectrum of $H$ that is off the real axis leads to a dynamical instability. Hence we say that the system is {\em dynamically stable} if the spectrum of the BdG Hamiltonian $H$ lies on the real axis. The following simple result states that thermodynamic stability in the sense of Section~\ref{sec-QuadraticHam} leads to dynamical stability.

\begin{proposi}
\label{prop-DynStab}
If $A=JH\geq 0$, then $\sigma(H)\subset\RM$. Furthermore, if $A$ is positive and $Q_\pm$ are Riesz projections on positive/negative parts of the spectrum, then
$$
\pm\,Q_\pm^*\,J\,Q_\pm
\;=\;
\pm\,J\,Q_\pm
\;\geq\;0
\;.
$$
\end{proposi}

\noindent {\bf Proof.} Consider the operator $G=A^{\frac{1}{2}}JA^{\frac{1}{2}}$. As $J$ is selfadjoint, also $G^*=G$. Hence $\sigma(G)\subset\RM$. On the other hand, by general principles $\sigma(G)\cup\{0\}=\sigma(JA^{\frac{1}{2}}A^{\frac{1}{2}})\cup\{0\}=\sigma(H)\cup\{0\}$. As to the second claim, the first identity follows from $Q_\pm^*=JQ_\pm J$ and the fact that $Q_\pm$ is idempotent. For the positivity, let us start from the Riesz formula along positively oriented paths $\Gamma_\pm$ in the right/left half-plane:
\begin{align*}
J\,Q_\pm
& \;=\;
\oint_{\Gamma_\pm}\frac{dz}{2\pi\imath}\;(z\,J-A)^{-1}
\\ &
\;=\;
(A^{-\frac{1}{2}})^*
\Big(\oint_{\Gamma_\pm}\frac{dz}{2\pi\imath}\;(z\,G^{-1}-\one)^{-1}\Big)
A^{-\frac{1}{2}}
\\ &
\;=\;
(A^{-\frac{1}{2}})^*
\Big(\pm\,\oint_{-\Gamma_\pm}\frac{d\xi}{2\pi\imath}\;(\xi\,\one-G^{-1})^{-1}\Big)
A^{-\frac{1}{2}}
\;,
\end{align*}
where $-\Gamma_\pm$ is the inverted path, but positively oriented. Hence the integral is the orthogonal spectral projection of the selfadjoint operator $G^{-1}$. This shows the claim.
\hfill $\Box$

\vspace{.2cm}

Let us note that the proof only used the $J$-selfajointness \eqref{eq-JHermition} of $H$, and not the PHS \eqref{eq-BdGsymmetry}. It also ought to be stressed that the inverse implication to Proposition~\ref{prop-DynStab} is not true. Examples of BdG Hamiltonians with real spectrum, but no sign for $JH$ will be provided shortly. Finally let us comment on the positivity of $\pm J Q_\pm$. If $Q_+$ is the finite dimensional projection associated to a discrete eigenvalue $\lambda>0$ of $H$, then the positivity can be restated as saying that $\lambda$ is $J$-definite and its Krein signature is positive. Negative eigenvalues of $H$ are also $J$-definite with negative Krein signature (see, {\it e.g.}, \cite{Lan,SB3} for a definition of these notions). More generally, the whole positive spectrum of $H$ is positive $J$-definite, and the negative spectrum is negative $J$-definite. By Krein stability theory ({\it e.g.} \cite{SB3,SV}), these facts imply that the spectrum of $H$ with $JH\geq 0$ can leave the real axis only via the origin, namely when $A$ ceases to be invertible. This is, of course, in agreement with the first claim of Proposition~\ref{prop-DynStab}.

\vspace{.2cm}

Now let us analyze how varying the chemical potential $\mu\in\RM\mapsto H_\mu$ in \eqref{eq-JHermitianMu} may either help to establish dynamical stability or destroy it. This will also illustrate the generic Krein collisions. As already pointed out, if $\mu$ is smaller than $-\|A\|$, clearly $H_\mu$ is thermodynamically and thus also dynamically stable. As $\mu$ is increased, the invertibility of $H_\mu$ may be lost and the spectra of different Krein signature collide in $0$. Generically, the spectrum then leaves the real axis via a tangent bifurcation. Let us illustrate this by the $2\times 2$  BdG Hamiltonian
$$
H_{\mu,\nu}\;=\;
\begin{pmatrix}
\mu & \imath\,\nu \\ \imath\,\nu & -\mu
\end{pmatrix}
\;,
\qquad
\nu\in\RM
\;.
$$
Its eigenvalues are given by $\pm\sqrt{\mu^2-\nu^2}$ and for $\nu\not=0$, the spectrum leaves the real axis at $\mu=-\nu$, and then joins it again at $\mu=\nu$. This is the generic scenario also for a BdG Hamiltonian on an infinite Hilbert space, in particular, whenever $\Delta\not=0$. This is how dynamically unstable edge modes will arise in Section~\ref{sec-UnstableBound}. For sake of completeness let us also present a $4\times 4$ BdG Hamiltonian as a toy model for the other generic Krein collision allowing eigenvalues to leave off the real axis, namely a quadrouple collision:
$$
H(\lambda,\nu)
\;=\;
\begin{pmatrix}
\lambda & 0 & 0 & \imath\,\nu
\\
0 & - \lambda & \imath\,\nu & 0
\\
0 & \imath\,\nu & -\lambda & 0 
\\
\imath\,\nu & 0 & 0 & \lambda
\end{pmatrix}
\;,
\qquad
\lambda\,,\,\nu\,>\,0
\;.
$$
The $4$ eigenvalues are $\pm\lambda\pm\imath\nu$ and move off the real axis while respecting the spectral symmetry \eqref{eq-FourFold}. A notable (non-generic) exception to these instabilities is the path $\mu\mapsto H_\mu$ when $\Delta=0$. Then $H_\mu=\diag(h-\mu,-(\overline{h}-\mu))$ and the spectra of opposite signature overlap without leaving the real axis. This is used in Section~\ref{sec-UnstableBound} to produce an inversion of Krein signatures.

\subsection{Bogoliubov transformation}
\label{sec-Bogo}

A Bogoliubov transformation is an invertible linear operator $T$ on $\Hh_\ph$ changing the BdG Hamiltonian $H$ to $THT^{-1}$ while conserving all structural properties of \eqref{eq-HBdG3}, namely after the basis change in
$$
{\bf H}
\;=\;
\frac{1}{2}\;
\begin{pmatrix} \crea \\ -\anni \end{pmatrix}^\trans 
T^{-1} \big(TH T^{-1} \big)
T\begin{pmatrix} \anni \\ \crea \end{pmatrix}
\;,
$$
one has $B=THT^{-1}\in \HM(\Hh_\ph,J,K)$ and the entries of the vector
$$
\begin{pmatrix} \mathfrak{b} \\ \mathfrak{b}^* \end{pmatrix}
\;=\;
T\begin{pmatrix} \anni \\ \crea \end{pmatrix}
$$
satisfy the CCR's as well as
$$
\begin{pmatrix} \mathfrak{b}^* \\ - \mathfrak{b} \end{pmatrix}^\trans 
\;=\;
\begin{pmatrix} \crea \\ - \anni \end{pmatrix}^\trans 
T^{-1}
\;.
$$
All of this is guaranteed if $T\in\UM(\Hh_\ph,J,K)$. Indeed, then
$$
JB^*J\;=\;J(T^{-1})^*JHJT^*J\;=\;THT^{-1}\;=\;B\;,
\qquad
K\overline{B}K
\;=\;
K\overline{T}KHK(\overline{T})^{-1}K
\;=\;
B
\;.
$$
Furthermore, the CCR's for $\mathfrak{b}$ and $\mathfrak{b}^*$ follow from \eqref{eq-CCR} combined with $T^\trans IT=I$,
and finally using $T^{-1}=I^*T^\trans I$
$$
\begin{pmatrix} \crea \\ - \anni \end{pmatrix}^\trans 
T^{-1}
\;=\;
\begin{pmatrix} \anni \\ \crea  \end{pmatrix}^\trans 
T^\trans I
\;=\;
\left[T\begin{pmatrix} \anni \\ \crea \end{pmatrix}\right]^\trans I
\;=\;
\begin{pmatrix} \mathfrak{b} \\ \mathfrak{b}^* \end{pmatrix}^\trans I
\;=\;
\begin{pmatrix} \mathfrak{b}^* \\ - \mathfrak{b} \end{pmatrix}^\trans 
\;.
$$
Hence, one can write
$$
{\bf H}
\;=\;
\frac{1}{2}\;
\begin{pmatrix} \mathfrak{b}^* \\ -\mathfrak{b} \end{pmatrix}^\trans  B
\begin{pmatrix} \mathfrak{b} \\ \mathfrak{b}^* \end{pmatrix}
$$
Of course, the aim is to choose $T\in\UM(\Hh_\ph, J,K)$ such that $B$ is particularly simple. In particular, diagonal $B$ is possible if $JH$ is strictly positive (so, in particular, thermodynamically stable) as shows the following well-known result, {\it e.g.} \cite{BR}.

\begin{proposi}
\label{prop-diag}
If $A=JH>0$, then exists $T\in\UM(\Hh_\ph,J,K)$ such that $D=T HT^{-1}\in \HM(\Hh_\ph,J,K)$ is diagonal, selfadjoint and $JD>0$.
\end{proposi}

\noindent {\bf Proof.} One has
$$
H
\;=\;
JA
\;=\;
(A^{\frac{1}{2}})^{-1}(A^{\frac{1}{2}}JA^{\frac{1}{2}})A^{\frac{1}{2}}
\;=\;
(A^{\frac{1}{2}})^{-1}GA^{\frac{1}{2}}
\;,
$$
where we set 
$$
G\;=\;
A^{\frac{1}{2}}\,J\,A^{\frac{1}{2}}
\;=\;
(JH)^{\frac{1}{2}}\,J\,(JH)^{\frac{1}{2}}
\;.
$$
Due to \eqref{eq-BdGsymmetry0} and $JK=-KJ$, one then has
$$
G\;=\;G^*
\;,
\qquad
K\,\overline{G}\,K\;=\;-G
\;.
$$
By functional calculus of selfadjoint operators, there exists a unitary $U$ on $\Hh_\ph$ satisfying $K\overline{U}K=U$ such that $UGU^*=D$ is diagonal and $D=\diag(d,-d)$ with $d>0$. Then set $T=|D|^{-\frac{1}{2}}UA^{\frac{1}{2}}$. One has
$$
T^*JT
\;=\;
A^{\frac{1}{2}}U^*
|D|^{-\frac{1}{2}}J
|D|^{-\frac{1}{2}}UA^{\frac{1}{2}}
\;=\;
A^{\frac{1}{2}}U^*D^{-1}UA^{\frac{1}{2}}
\;=\;
A^{\frac{1}{2}}G^{-1}A^{\frac{1}{2}}
\;=\;
J
\;,
$$
and
$$
K\overline{T}K
\;=\;
\big(K|D|^{\frac{1}{2}}K\big)\,\big(K\overline{U}K\big)\,\big(K\overline{A^{\frac{1}{2}}}K\big)
\;=\
|D|^{\frac{1}{2}}\,U\,A^{\frac{1}{2}}
\;=\;
T
\;.
$$
Hence $T\in\UM(\Hh_\ph,J,K)$. Finally
$$
T HT^{-1}
\;=\;
|D|^{-\frac{1}{2}}UA^{\frac{1}{2}}JA^{\frac{1}{2}}U^*|D|^{\frac{1}{2}}
\;=\;
|D|^{-\frac{1}{2}}D|D|^{\frac{1}{2}}
\;=\;
D
\;,
$$
which completes the proof.
\hfill $\Box$

\vspace{.2cm}

Recent results \cite{BB,NNS} have shown that the conditions in Proposition~\ref{prop-diag} can be considerably relaxed. In particular, strict positivity of $A$ is not needed. For the present purposes the above form is sufficient. One of the main implications of Proposition~\ref{prop-diag} is that it allows to define functional calculus of $H=JA$ with $A>0$ by
$$
f(H)
\;=\;
T^{-1}\,f(D)\,T
\;,
$$
for any measurable function $f:\RM\to\CM$. For a real valued function $f$, one has 
$$
f(H)^*
\;=\;
Jf(H)J
\;,
\qquad
K^*f(\overline{H})K\;=\;f(-H)
\;.
$$
Functional calculus can actually be defined for the larger class of so-called definitizable operators \cite{Lan} which also also have local $J$-definite spectrum, but intervals of opposite $J$-definitivity can alternate a finite number of times.

\vspace{.2cm}

A natural question is if a Bogoliubov transformation $T$ is implementable in Fock space, namely whether there is a unitary $U_T$ on $\Ff$ such that
$$
T\begin{pmatrix} \anni \\ \crea \end{pmatrix}
\;=\;
U_T
\begin{pmatrix} \anni \\ \crea \end{pmatrix}
U_T^*
\;.
$$
This implementability is characterized in a celebrated theorem by Shale, see again \cite{BB,NNS}. In situations of interest to solid state physics, it is not given.

\section{Topological invariants of BdG Hamiltonians}
\label{sec-TopInvBdG}

In this section,  the Hamiltonian is supposed to be a covariant family of short-range BdG Hamiltonians $H=(H_\omega)_{\omega\in\Omega}$ on $\Hh_\ph=\ell^2(\ZM^d)\otimes\CM^{2L}$ each satisfying \eqref{eq-BdGsymmetry} and \eqref{eq-JHermition}. All short-range covariant operator families form a C$^*$-algebra $\Aa$ and the hypothesis is thus simply that $H\in\Aa$ and \eqref{eq-BdGsymmetry} and \eqref{eq-JHermition} hold. For the convenience of the reader and to fix notations, the essentials on $\Aa$ and operators therein a collected in the Appendix. The second main assumption is that $H$ is dynamically stable, namely $\sigma(H_\omega)\subset\RM$ for all $\omega\in\Omega$. Moreover, the positive spectrum is supposed to be contained in disjoint intervals 
$$
0\,<\,I_1\,<\,I_2\,<\,\ldots\,<\,I_J
\;.
$$   
Due to \eqref{eq-FourFold}, one then knows that the negative spectrum of $H$ is contained in the intervals $I_{-j}=-I_j$. The associated spectral Riesz  projections are denoted by
$$
Q_j
\;=\;
\chi_{I_j}(H)
\;\in\;\Aa
\;,
\qquad
j=-J,\ldots,J
\;.
$$
Note that these idempotents satisfy \eqref{eq-Q*}, namely
\begin{equation}
\label{eq-P*}
Q_j^*
\;=\;
J\,Q_j\,J
\;,
\qquad
K\,\overline{Q_j}\,K
\;=\;
Q_{-j}
\;.
\end{equation}
General principles allow to extract topological numbers from these projections $Q_j\in\Aa$, more precisely (higher) Chern numbers and, if the system has a chiral symmetry, (higher) winding numbers which are sometimes also called odd Chern numbers \cite{PS}. The standard theory, see {\it e.g.} \cite{PS}, addresses only orthogonal projections so that some additional care is needed when defining these objects, as was already noted in \cite{SMMO}. Instead of repeating the general definitions from \cite{PS}, we focus here on the usual (first) Chern numbers, say in the $1$ and $2$-direction (hence $d\geq 2$). They are defined as in \cite{Bel,PS} by
$$
\Ch(Q)
\;=\;
2\pi\imath\;
\Tt(Q[\nabla_1Q,\nabla_2 Q])
\;,
$$
where $\Tt$ is the trace per unit volume and $\nabla_1$ and $\nabla_2$ are non-commutative derivatives in the two spatial dimensions. They replace integration over the Brillouin zone and derivations w.r.t. quasimomenta respectively in disordered systems. Detailed definitions are recalled in the Appendix. For an orthogonal projection in $\Aa$ and for dimension $d=2,3$, the above Chern number is known to be integer-valued and equal to the index of a suitable Fredholm operator \cite{Bel,PS}. This result extends to non-selfadjoint projections. 

\begin{proposi}
\label{prop-ChernQuant}
Let $Q=(Q_\omega)_{\omega\in\Omega}\in\Aa$ be a covariant family of idempotents in the crossed product algebra $\Aa$.  The Chern number $\Ch(Q)$ is a homotopy invariant (under homotopies of idempotents in $\Aa$) which is equal to an integer for $d=2,3$.
\end{proposi}

\noindent {\bf Proof.} We will show that $Q$ can be homotopically deformed into an orthogonal projection by a differentiable path $t\in[0,1]\mapsto Q(t)$ of idempotents with $Q(0)=Q$ and $Q(1)^*=Q(1)$, and that along the path $\Ch(Q(t))$ does not change. Let us choose $Q(1)$ to be the (orthogonal) range projection of $Q$. There are numerous formulas for it, for example
$$
Q(1)\;=\;Q\;\big(\one+(Q-Q^*)\big)^{-1}\;\in\;\Aa
\;.
$$
Note that the inverse exists because the spectrum of $Q-Q^*$ is purely imaginary. Then 
$$
Q(t)\;=\;Q(1)\,+\,t(Q-Q(1))
\;,
$$
is a path with the above properties. In particular, $Q(1)Q=Q$ implies that $Q(t)$ is an idempotent. By the Leibniz rule, $\partial_t\Ch(Q(t))$ contains three summands. As 
$$
\partial_t Q(t)
\;=\;
Q(t)\partial_t Q(t)(\one-Q(t))\;+\;(\one-Q(t))\partial_t Q(t) Q(t)
\;
$$  
and similarly for $\nabla_j Q(t)$, the cyclicity of $\Tt$  implies that all terms of  $\partial_t\Ch(Q(t))$ cancel out.
\hfill $\Box$

\vspace{.2cm}

Before going on, let us note a few supplementary facts about the Chern numbers already spelled out in \cite{SMMO}.

\begin{proposi}
\label{prop-ChernProperties}
Suppose that $H$ is thermodynamically stable. Then one has
$$
\Ch(Q_j)\;=\;-\,\Ch(Q_{-j})
\;,
$$
as well as the sum rule for all positive energy bands
$$
\sum_{j=1}^J\,\Ch(Q_j)
\;=\;
0
\;.
$$
\end{proposi}

\noindent {\bf Proof.} The first claim follows from the second identity in \eqref{eq-P*} when the definition of $\Ch(Q_j)$ is complex conjugated and the invariance of $\Tt$ under $K$ is used. For the second identity, one can use the homotopy $H(t)=J((1-t)A+t\one)$ of thermodynamically stable BdG Hamiltonians. As $0\not\in\sigma(H(t))$ for all $t$, the positive spectral projection $Q(t)$ is well-defined for all $t$. For $H(1)=J$ one has $Q(1)=\frac{1}{2}(J+\one)$ with $\Ch(Q(1))=0$. On the other hand, the additivity of the Chern numbers implies that $\Ch(Q(0))=\sum_{j=1}^J\,\Ch(Q_j)$.
\hfill $\Box$

\vspace{.2cm}

A simple way to produce examples of topologically non-trivial models is to start out with a well-known topological one-particle Hamiltonian $h$. It's topology is then conserved under the addition of a weak pairing term $\Delta$. This procedure is applied in Section~\ref{sec-UnstableBound}. There, however, also situations where the pairing operator $\Delta$ is crucial for opening gaps and producing topological bands. In \cite{PHB}, $h$ is chosen as the discrete Laplacian on the kagom\'e lattice and then a suitable parametric driving is added. The most simple instance considered in \cite{PHB} is
$$
h
\;=\;
\begin{pmatrix}
0 & \one+S_1^* & \one +S_2^* \\
\one+S_1 & 0 & \one +S_3^* \\
\one+S_2 & \one +S_3 & 0
\end{pmatrix}
\;,
\qquad
\Delta
\;=\;
\nu
\begin{pmatrix}
\one & 0 & 0 \\
0 & e^{\frac{2\pi\imath}{3}}\,\one & 0 \\
0 & 0 & e^{\frac{4\pi\imath}{3}}
\end{pmatrix}
\;,
$$
where $S_3=S_2^*S_1$. For $\mu$ sufficiently large and $\nu$ small, $H_\mu$ given by \eqref{eq-JHermitianMu} with \eqref{eq-HBdG3} has three positive bands with Chern numbers equal to $-1$, $0$ and $1$ respectively.

\vspace{.2cm}

Now let us comment on the definition of so-called odd Chern numbers \cite{PS} for covariant Hamiltonians having a chiral symmetry. This symmetry typically results from a bipartite structure of the model. On the one-particle Hilbert space $\Hh$ it shall be implemented by a real selfadjoint unitary $L$ which (possibly after an adequate basis change) is of the form $L=\diag(\one,-\one)$ w.r.t. to a given grading $\Hh=\Hh_+\oplus\Hh_-$. This operator is diagonally extended to $\Hh_\ph=\Hh\otimes\CM^2_\ph$. The chiral symmetry of the BdG Hamiltonian then reads
$$
L\,H\,L
\;=\;
-\,H
\;.
$$
In the grading $\Hh_\ph=(\Hh_+\otimes \CM^2_\ph)\oplus(\Hh_-\otimes \CM^2_\ph)$, this implies that
$$
H
\;=\;
\begin{pmatrix} 0 & B \\ JB^*J & 0 \end{pmatrix}
\;,
$$
where $B\in\Aa$ is supposed to be invertible, or equivalently that $H$ is invertible. From $B$ one can now extract the odd Chern numbers. The one of lowest degree is the winding number (say in the $1$-direction)
$$
\mbox{\rm Wind}(B)
\;=\;
\imath\;\Tt(B^{-1}\nabla_1 B)
\;.
$$
Let us note that it is suffienct that the chiral symmetry holds approximately for these invariants to be well-defined \cite{PS}. In dimension $d=1,2$ it is known that $\mbox{\rm Wind}(B)$ is an integer \cite{PS}. It is straightforward to write out variants of the SSH model to produce examples of chiral BdG Hamiltonians in dimensions $d=1$ which have non-trivial $\mbox{\rm Wind}(B)$. Higher winding numbers (that is, higher odd Chern numbers) are defined as in \cite{PS}. 

\section{Edge states of stable two-dimensional systems}
\label{sec-EdgeThermoStab}

In this section, $d=2$ and the bulk BdG Hamiltonian is supposed to be of the form $H=JA$ with $A>0$, namely the system is thermodynamically stable. As already stressed in the introduction, it is now of interest to show how the non-trivial Chern numbers imply the existence of edge state bands. Hence let us restrict the system to a half-space. The half-space BdG Hamiltonian $\widehat{H}=(\widehat{H}_\omega)_{\omega\in\Omega}$ acts on the half-space particle-hole Hilbert space $\widehat{\Hh}_\ph=\ell^2(\ZM\times\NM)\otimes\CM^{2L}$. Similar to \cite{KRS,PS}, these operators are of the form
\begin{equation}
\label{eq-HalfSpaceHam}
\widehat{H}_\omega
\;=\;
\Pi\, H_\omega\,\Pi^*\;+\;J\,\widetilde{A}_\omega
\;=\;
J\big(\widehat{A}_\omega\;+\;\widetilde{A}_\omega\big)
\;,
\end{equation}
where $\Pi:\ell^2(\ZM^2)\otimes\CM^{2L}\to \ell^2(\ZM\times\NM)\otimes\CM^{2L}$ is the partial isometry given by restriction of wave functions and $\widehat{A}_\omega= \Pi{A}_\omega\Pi^*$, and finally $\widetilde{A}_\omega\geq 0$ is norm limit of operators acting only on a strip $\ZM\times\{1,\ldots,N\}$ and being covariant in the infinite direction, that is $S_1\widetilde{A}_\omega S_1^*=\widetilde{A}_{T_1\omega}$. In the terminology of \cite{KRS,PS}, $\widetilde{A}$ lies in the ideal $\Ee$ of edge operators sitting inside the Toeplitz extension $T(\Aa)$ of half-space operators, see again the Appendix or \cite{PS}. Associated to $\widehat{H}$ is then a quadratic operator $\widehat{\HH}$ on the bosonic Fock space over $\ell^2(\ZM\times\NM)\otimes\CM^{L}$, similar as in \eqref{eq-HalfFock}. 

\vspace{.2cm}

The first important fact is that also $\widehat{H}$ is thermodynamically stable by construction. Due to Proposition~\ref{prop-DynStab} this implies that $\sigma(\widehat{H})\subset\RM$ and thus there are no dynamically unstable boundary states as in the situations described in Section~\ref{sec-UnstableBound}. Moreover, as $\widehat{H}$ is invertible there are no edge states with zero energy either. However, the following result shows that there are bands of edge states between $I_j$ and $I_{j+1}$ if $\Ch(Q_j)\not=\Ch(Q_{j+1})$. Moreover, these bands lead to a quantized matter current density along the boundary. This density is calculated using  $\widehat{\Tt}$, a combination of the the trace per unit volume along the boundary and the usual trace in the direction perpendicular to the boundary. This is a densely defined tracial state on $\Ee$, and a formula is given in the Appendix.

\begin{theo}
\label{theo-StableBoundary}
Suppose given the set-up described in this section. Let $g_j:\RM\to[0,\infty)$ be a smooth function of unit integral and supported in the gap between $I_j$ and $I_{j+1}$ for $j=1,\ldots,J-1$. Then $g_j(\widehat{H})$ is $\widehat{\Tt}$-traceclass and
$$
\frac{1}{2\pi}\;\widehat{\Tt}\big(g_j(\widehat{H})\nabla_1 \widehat{H}\big)
\;=\;
-\;
\sum_{i=1}^j\Ch(Q_i)
\;.
$$ 
In particular, if the integer on the r.h.s. does not vanish, then the gap between $I_j$ and $I_{j+1}$ lies in the spectrum of $\widehat{H}$. 
\end{theo}

This theorem will be proved by an application of the fermionic theory from \cite{KRS,PS} to the selfadjoint operator
$$
G\;=\;A^{\frac{1}{2}}\,J\,A^{\frac{1}{2}}
\;.
$$
This operator was already used in the proof of Proposition~\ref{prop-diag} where it was shown that there exists an invertible linear operator $M$ (equal to $U^*T$ in the notations of that proof) such hat 
$$
M\,H\,M^{-1}\;=\;G
\;.
$$
In particular, the spectra of $H$ and $G$ coincide.  Hence it is also natural to introduce the orthogonal projections
$$
P_j\;=\;\chi_{I_j}(G)
\;.
$$
Furthermore, the construction in  Proposition~\ref{prop-diag} shows that $M\in\Aa$ is a covariant operator family, and thus also $G$ and the $P_j$ are in $\Aa$. 

\begin{proposi}
\label{prop-ChernLink}
For all $j=-J,\ldots,J$, 
$$
\Ch(P_j)\;=\;\Ch(Q_{j})
\;.
$$
\end{proposi}

\noindent {\bf Proof.} By functional calculus $P_j= MQ_jM^{-1}$. The polar decomposition $M=|M|V$ with unitary $V$ allows to define a differentiable homotopy of idempotents $t\in[0,1]\mapsto |M|^tVQ_jV^*|M|^{-t}$ connecting $P_j$ to $VQ_jV^*$. By the argument in the proof of Proposition~\ref{prop-ChernProperties}, one has 
$$
\Ch(P_j)
\;=\;
\Ch(VQ_{j}V^*)
\;=\;
\Ch\big(\diag(V,V^*)\,\diag(Q_j,0)\,\diag(V,V^*)^*\big)
\;=\;
\Ch(Q_j)
\;,
$$
where in the last equality the standard path from $\diag(V,V^*)$ to the identity is used.
\hfill $\Box$

\vspace{.2cm}

\noindent {\bf Proof} of Theorem~\ref{theo-StableBoundary}:  Let $\widehat{G}=\Pi G\Pi^*+\widetilde{G}$ be of the form \eqref{eq-HalfSpaceHam} with an operator $\widetilde{G}$ lying in the edge algebra $\Ee$. For a smooth function $g_j$ supported by a gap of  $G$, it is known that $g_j(\widehat{G})$ is $\widehat{\Tt}$-traceclass \cite{PS}. The bulk-boundary correspondence \cite{KRS,PS} for the selfadjoint and gapped Hamiltonian $G$ states that 
\begin{equation}
\label{eq-HermitianBBC}
\sum_{i=-J}^j\Ch(P_i)
\;=\;
-\;\frac{1}{2\pi}\;\widehat{\Tt}\big(g_j(\widehat{G})\nabla_1 \widehat{G}\big)
\;.
\end{equation}
Let us stress that the identity \eqref{eq-HermitianBBC} is independent of the choice of the boundary operator $\widetilde{G}$. We will choose $\widetilde{G}$ such that
\begin{equation}
\label{eq-Gchoice}
\widehat{G}
\;=\;
\widehat{A}^{\frac{1}{2}}\,J\,\widehat{A}^{\frac{1}{2}}
\;,
\end{equation}
where $\widehat{A}=\Pi A\Pi^*$. This is possible because for any covariant operator families $B,C\in\Aa$, one knows that $\Pi BC\Pi^*-\Pi B\Pi^*\Pi C\Pi^*$ is in the edge algebra. Hence $\Pi A^{\frac{1}{2}}JA^{\frac{1}{2}}\Pi^*-\Pi A^{\frac{1}{2}} \Pi^* J \Pi A^{\frac{1}{2}}\Pi^*$ is in the edge algebra. Moreover, for any smooth real-valued function $f:\RM\to\RM$ one knows that $\Pi f(A)\Pi^*-f(\Pi A\Pi^*)$ is $\widehat{\Tt}$-traceclass \cite{PS}. Now the square root is not smooth, but because $A>0$ one can modify it to a smooth function $f$ so that $A^{\frac{1}{2}}=f(A)$. In conclusion, one can indeed choose $\widetilde{G}$ such that \eqref{eq-Gchoice} holds. Invoking Propositions~\ref{prop-ChernProperties} and \ref{prop-ChernLink}, we conclude that
$$
\sum_{i=1}^j\Ch(Q_i)
\;=\;
-\;\frac{1}{2\pi}\;\widehat{\Tt}\big(g_j(\widehat{G})\nabla_1 \widehat{G}\big)
\;.
$$
If the l.h.s. does not vanish, then $\widehat{G}$ must have spectrum throughout the gap between $I_j$ and $I_{j+1}$. The same then holds for $\widehat{H}$ because $\sigma(\widehat{G})=\sigma(\widehat{H})$, {\it cf.} the proof of Proposition~\ref{prop-DynStab}. Furthermore, as in that proof, there exists a half-space operator $\widehat{M}$ such that  $\widehat{M}\widehat{H}\widehat{M}^{-1}=\widehat{G}$ and again $\widehat{M}\in T(\Aa)$.  Hence
\begin{align*}
\widehat{\Tt}\big(g_j(\widehat{G})\nabla_1 \widehat{G}\big)
&
\;=\;
\widehat{\Tt}\big(\widehat{M}^{-1}g_j(\widehat{H})\widehat{M}\;\nabla_1(\widehat{M}^{-1} \widehat{H}\widehat{M})\big)
\\
&
\;=\;
\widehat{\Tt}\big(g_j(\widehat{H})\widehat{M}\;
(\nabla_1 \widehat{M}^{-1}\, \widehat{H}\,\widehat{M}
\,+\,
\widehat{M}^{-1}\, \nabla_1\widehat{H}\,\widehat{M}
\,+\,
\widehat{M}^{-1}\, \widehat{H}\,\nabla_1\widehat{M})
\;
\widehat{M}^{-1}\big)
\\
&
\;=\;
\widehat{\Tt}\big(g_j(\widehat{H})\;
\nabla_1\widehat{H}
\big)
\;,
\end{align*}
concluding the proof.
\hfill $\Box$

\section{Dynamically unstable bound states}
\label{sec-UnstableBound}

The aim of this section is to show how selected bound states of a bosonic quadratic Hamiltonian can become dynamically unstable in the presence of  parametric driving or in a Bose-Einstein condensate after a quench. The basic idea seems to go back to Barnett \cite{Bar} who considered a particular one-dimensional situation.  While the instability is merely a special case of the tangent bifurcation described in Section~\ref{sec-KreinPer}, our main interest is in the dynamical instability of topological bound states in an extended system having dynamically stable bulk eigenstates. Below, two different scenarios are discussed where the stability of the bulk  is robust in presence of disorder. In the first scenario, the dynamical stability of the bulk BdG Hamiltonian is protected by a symmetry for any disorder that does not break that symmetry. In the second scenario, the dynamical stability is protected  for any arbitrary type of disorder by band gaps separating bands with different Krein signatures. 

\vspace{.2cm}

In the first scenario, the generic setup consists of a one-particle real Hamiltonian $h=\bar{h}=h^t$ on a Hilbert space $\Hh=\ell^2(\ZM^d)\otimes\CM^{2L}$ as well as a modified Hamiltonian $\widehat{h}=\widehat{h}^*$ on a half-space Hilbert space $\widehat{\Hh}=\ell^2(\ZM^{d-1}\times\NM)\otimes\CM^{2L}$. As in Section~\ref{sec-EdgeThermoStab}, we refer to $h$ as the bulk one-particle Hamiltonian and $\widehat{h}$ as the half-space one-particle Hamiltonian. Associated with a given chemical potential $\mu\in\RM$ and a driving amplitude $\nu\in\RM$, let us now consider BdG Hamiltonians on the particle-hole Hilbert spaces $\Hh_\ph$ and $\widehat{\Hh}_\ph$:
\begin{equation}
\label{eq-HamParDri}
H_{\mu,\nu}
\;=\;
\begin{pmatrix}
h-\mu & \imath\,\nu \\
\imath\,\nu & -(\overline{h}-\mu)
\end{pmatrix}
\;,
\qquad
\widehat{H}_{\mu,\nu}
\;=\;
\begin{pmatrix}
\widehat{h}-\mu & \imath\,\nu \\
\imath\,\nu & -(\overline{\widehat{h}}-\mu)
\end{pmatrix}
\;.
\end{equation}
Here $\mu$ and $\nu$ are seen as multiples of the identity operator. Via \eqref{eq-HBdG3} one can now associate (formal) operators on Fock space, namely up to a constant term:
$$
\HH_{\mu,\nu}
\;=\;
(\crea)^\trans \,h\,\anni\;-\;\mu\,{\bf N}
\;+\;
\frac{\imath\,\nu}{2}\;\big( (\crea)^\trans\,\crea\,-\,\anni^\trans \,\anni\big)
\;.
$$
(Recall that ${\bf N}$ is defined as in \eqref{eq-NumberOp}.) Similarly, with $\hat{\anni}$ and $\hat{\anni}^*$ denoting the vectors of annihilation and creation operators over the half-space Hilbert space,
\begin{equation}
\label{eq-HalfFock}
\widehat{\HH}_{\mu,\nu}
\;=\;
(\hat{\anni}^*)^\trans \,\widehat{h}\,\hat{\anni}\;-\;\mu\,\widehat{{\bf N}}
\;+\;
\frac{\imath\,\nu}{2}\;\big( (\hat{\anni}^*)^\trans\,\hat{\anni}^*\,-\,\hat{\anni}^\trans \,\hat{\anni}\big)
\;.
\end{equation}
The factors involving $\nu$ are  parametric driving terms (in the rotating frame) or, for a Bose-Einstein condensate after a quench,   describe the interactions treated at the mean field level. A particular example of this scenario is discussed in \cite{Bar} where   $h$ is the  SSH model Hamiltonian.

\vspace{.2cm}

The physically interesting aspect of the following proposition is that there are parameter regimes for which the bound states become dynamically unstable while the bulk modes remain dynamically stable. This is of interest in $d=1$ when $h$ is a possibly disordered chiral Hamiltonian having a non-trivial winding number. Its half-space restriction $\widehat{h}$ then has a topologically protected zero energy bound state which generically is of the same multiplicity as the winding number, see Chapter~1 in \cite{PS}.  This zero mode belongs to the discrete spectrum of $\widehat{h}$, but does not lie  in the spectrum of the initial operator $0\not\in\sigma(h)$. As to the essential spectra, one has $\sigma_\ess(\widehat{h})=\sigma_\ess(h)$.

\begin{proposi}
\label{prop-DynUnstable} Suppose that $h=\overline{h}$ is real. The spectrum of $\widehat{H}_{\mu,\nu}$ is given
$$
\sigma(\widehat{H}_{\mu,\nu})
\;=\;
-\,\left(\sigma(\widehat{h}-\mu)^2-\nu^2\right)^{\frac{1}{2}}\;\cup\;
\left(\sigma(\widehat{h}-\mu)^2-\nu^2\right)^{\frac{1}{2}}
\;,
$$
and similarly for the spectrum of $H_{\mu,\nu}$. For $|\mu|<|\nu|<g-|\mu|$, the bulk Hamiltonian $H_{\mu,\nu}$ is dynamically stable and the bound states of the Hamiltonian $\widehat{H}_{\mu,\nu}$ are dynamically unstable. 
\end{proposi}

\noindent {\bf Proof.} Due to the reality $h=\overline{h}$, the Hamiltonian $\widehat{H}_{\mu,\nu}$ can be diagonalized by a diagonal unitary of the form $U=u\oplus u$. Let us go to this spectral representation and let $e$ be a possibly generalized eigenvalue of $\widehat{h}$. Then the eigenvalues of the $2\times 2$ matrix
$$
\begin{pmatrix} e-\mu & \imath\,\nu \\ \imath\,\nu & -(e-\mu) \end{pmatrix} 
$$
are in the spectrum of $\widehat{H}_{\mu,\nu}$. They are given by $E_\pm(\mu,\nu)=\pm\big((e-\mu)^2-\nu^2\big)^{\frac{1}{2}}$. From this one can read off all claims. In particular, for the zero mode $e=0$ of $\widehat{h}$, the eigenvalues $E_\pm(\mu,\nu)=\pm\big(\mu^2-\nu^2\big)^{\frac{1}{2}}$ undergo a Krein collision at $\mu=\nu$. For the spectrum of $H_{\mu,\nu}$ one argues in the same manner.
\hfill $\Box$

\vspace{.2cm}

Let us stress that in the scenario discussed above the stability of the bulk is protected for any non-magnetic disorder (disorder that does not violates the reality condition $h=\bar{h}$). Without the reality assumption, the spectrum generically undergoes  quadrouple Krein collisions if  bands with opposite Krein signature overlap. This is usually the case because if the half-space restriction $\widehat{h}-\mu$ has a zero eigenvalue, the full space operator $h-\mu$ is neither positive nor negative. Thus some fine tuning of $h-\mu$  is required to avoid the overlap, see below. There are other related scenarios where the single-particle Hamiltonian $h$ is not real but  the quadrouple collisions are nevertheless suppressed by a combination of various symmetries. One such case is analyzed in \cite{GLB} where $h$ is the Hamiltonian of a modified Kane and Mele model and the quadrouple collisions are suppressed by the combination of odd time reversal symmetry, inversion and spin rotation symmetry. Another important point in Proposition~\ref{prop-DynUnstable} is that in the absence of quadrouple Krein collisions the energies of the dynamically unstable edge states remain purely imaginary. As a consequence such excitations do not lead to any transport.

\vspace{.2cm}

Next, let us discuss the second scenario where a dynamical instability occurs for a topological bound state or a topological band  of edge states,  while the stability of the bulk is protected  by gaps separating bulk bands with opposite Krein signature. In this second scenario,  the Hamiltonians $H_{\mu,\nu}$ and $\widehat{H}_{\mu,\nu}$ are of the same form \eqref{eq-HamParDri}, but the single-particle Hamiltonian $h$ is not required to have a supplementary symmetry and the perturbation $\nu$ should be small, but need not be a multiple of the identity. On the other hand,  it is now crucial to choose  $h$ and the chemical potential $\mu$  such that  
\begin{equation}
\label{eq-BulkStableCond}
\sigma(h-\mu)\;\cap\;\big(-\sigma(\overline{h}-\mu)\big)
\;=\;\emptyset
\;,
\end{equation}
but at the same time
\begin{equation}
\label{eq-edgeunStableCond}
\sigma(\widehat{h}-\mu)\;\cap\;\big(-\sigma(\overline{\widehat{h}}-\mu)\big)
\;\neq\;\emptyset
\;.
\end{equation}
This assures that full space Hamiltonian $H_{\mu,0}$ has bands with definite Krein signature. As there are finitely many bands, this means that this operator is definitizable in the sense of Langer \cite{Lan}. In particular, the spectrum of $H_{\mu,0}$ is real even though $H_{\mu,0}$ need not be thermodynamically stable. If $\nu$ is small compared to the smallest band gap separating the bands of $\sigma({h}-\mu)$ and those of $-\sigma(\overline{h}-\mu)=-\sigma({h}-\mu)$, the full space Hamiltonian $H_{\mu,\nu}$ will remain dynamically stable. At the same time, if no symmetry suppresses the quadrouple Krein collisions for the topological states in the half-space system those states will become dynamically unstable. A generic setup that realizes this scenario consists of a two-dimensional Chern insulator described by a single-particle Hamiltonian $h$ whose lowest band $I'_1$ has non-zero Chern number and a bandwidth which is smaller that the width of the band gap separating it from the subsequent band $I'_2$. In this case, the chemical potential $\mu$ can be chosen such that
$$
I'_1-\mu\;<\;\mu\,-\,I'_1\;<\;I'_2-\mu\;.
$$
This  ensures that both conditions \eqref{eq-BulkStableCond} and \eqref{eq-edgeunStableCond} are fulfilled. In particular,  the interval $I''$ separating the intervals $I'_1-\mu$ and $\mu-I'_{1}$ represents a band gap separating bulk bands of opposite Krein signatures. On the other hand, the half-space reduction $\widehat{H}_{\mu,\nu=0}$ will have superposed bands of topological edge states  in $I''$ which have opposite Krein signatures. In the presence of a weak parametric interaction $\nu$, the full space BdG Hamiltonian  $H_{\mu,\nu}$  will remain stable while its half-space reduction will support in the interval $I''$ a band of dynamically unstable edge states. A specific implementation of this setup where the single-particle Hamiltonian $h$  is the Hofstadter model with rational magnetic flux (equal to $\frac{1}{4}$) has been  analyzed in detail in \cite{PHMC}.

\vspace{.2cm}

Let us emphasize that in most practical situations where certain symmetries might be only weakly broken it is useful to choose the appropriate form of $\nu$ to enhance the edge state instabilities without making the bulk unstable.   It is also worth to stress out that  in this second scenario the bands of dynamically unstable topological edge states  have complex energies and lead to transport along the physical boundary. This feature can be harnessed in useful topological devices like the tooplogical amplifier proposed in \cite{PHMC}.

\appendix

\section{Covariant operator families and their analysis}
\label{sec-alg}

This appendix reviews the C$^*$-algebraic description of covariant operators as introduced by Bellissard \cite{Bel} and described in \cite{PS}. This is a natural framework that allows to transpose well-known tools from Bloch theory for periodic operators, like derivation w.r.t. quasimomenta and integration over the Brillouin zone, to disordered or quasiperiodic systems. As constant magnetic fields are not compatible with genuine BdG operators ({\it e.g.} \cite{DS3}), only covariant operators w.r.t. the usual lattice translations (possibly with periodic or random magnetic fields) are covered. The formulation will be kept generic as in \cite{Bel,PS} and any addenda resulting from the BdG structure are described in the main text. Hence let us consider operators on a $d$-dimensional lattice Hilbert space $\ell^2(\ZM^d)\otimes\CM^M$ with $M$-dimensional fiber. In the main text, this will be the particle-hole Hilbert space so that $M=2L$. An orthonormal basis is given by $|n,m\rangle$ with $n=(n_1,\ldots,n_d)\in\ZM^d$ and $m=1,\ldots,M$. The shift operators and position operator are defined by
$$
S_j|n,m\rangle\;=\;|n+e_j,m\rangle
\;,
\qquad
X_j|n,m\rangle\;=\;n_j\,|n,m\rangle
\;,
$$
where $j=1,\ldots,d$ and $e_j$ is the $j$th standard basis vector. Given $a=(a_1,\ldots,a_d)\in\ZM^d$, let us also set
$$
S^a\;=\;S_1^{a_1}\cdots S_d^{a_d}
\;.
$$
Let now $(\Omega,T,\ZM^d,\PM)$ be a compact space equipped with a $\ZM^d$-action $T$ and an invariant and ergodic probability measure $\PM$. It is called the space  of disorder or crystalline configurations.  A finite-range covariant operator family  $A=(A_{\omega})_{\omega\in\Omega}$ consists of bounded operators on $\ell^2(\ZM^d)\otimes\CM^M$, strongly continuous in $\omega$, with $\langle n|A_\omega|n'\rangle=0$ for $|n-n'|>R$ for some finite $R$, and satisfying the covariance relation
$$
S^a
A_{\omega}
(S^a)^*
\;=\;
A_{T^a \omega}
\;,
\qquad a\in\ZM^d
\;.
$$
Now products and the adjoints of such families are again covariant operator families. A C$^*$-norm on the short-range covariant families is given by $\| A\| = \sup_{\omega \in \Omega}\| A_{\omega}\|$ and the corresponding closure is the reduced C$^*$-crossed product algebra $\Aa=C(\Omega)\rtimes\ZM^d$. On this algebra are defined unbounded and closed $*$-derivations $\nabla=(\nabla_1,\ldots,\nabla_d)$ by
\begin{equation}
\label{eq-derivrep}
(\nabla_jA)_\omega
\;=\;
\imath [A_{\omega}, X_j]
\;.
\end{equation}
Note that the r.h.s. is again formally covariant, but it is not in $\Aa$ for all covariant operator families. The common domain is denoted by $C^1 (\Aa)$. On it, the Leibniz rule holds
$$
\nabla (AB)\;=\; (\nabla A)B+A(\nabla B)
\;.
$$
Finally, a positive trace $\Tt$ on $\Aa$ is introduced by
\begin{equation}
\label{eq-tracepervol}
 \Tt(A) \; = \;
  \int_{\Omega}
    \PM(d\omega)\;
    \TR\bigl(\langle 0|A_\omega |0\rangle\bigr)
\;.
\end{equation}
Indeed, one has $\Tt(A^*A)\geq 0$ and $\Tt(A^*)=\overline{\Tt(A)}$ as well as $\Tt(AB)=\Tt(BA)$. Furthermore $\Tt(|A B|)\leq \|A\|\,\Tt(|B|)$ and $\Tt(\nabla A)=0$ for $A\in C^1(\Aa)$, so that also the partial integration $\Tt(A\nabla B)=-\Tt(\nabla A \,B)$ holds for $A,B\in C^1(\Aa)$. Moreover, Birkhoff's ergodic theorem implies that 
\begin{equation}
\label{eq-tracepervol2}
 \Tt(A) \; = \;
      \lim_{N\rightarrow\infty}
       \frac{1}{(2N+1)^d}
        \sum_{|n|\leq N}
         \TR\bigl(\langle n| A_{\omega} |n\rangle\bigr)
\mbox{ , }
\end{equation}
\noindent for $\PM$-almost all $\omega\in\Omega$. This shows that $\Tt$ is the trace
per unit volume.

\vspace{.2cm}

For the description of operators on the half-space Hilbert space $\ell^2(\ZM^{d-1}\times\NM)\otimes\CM^{M}$, we use as in \cite{KRS,PS} the Toeplitz extension $T(\Aa)$ of $\Aa$ given by operator families $\widehat{A}=(\widehat{A}_\omega)_{\omega\in\Omega}$ of the form
\begin{equation}
\label{eq-HalfSpaceOps}
\widehat{A}_\omega
\;=\;
\Pi\, A_\omega\,\Pi^*\;+\;\widetilde{A}_\omega
\;,
\end{equation}
where $\Pi:\ell^2(\ZM^d)\otimes\CM^{M}\to \ell^2(\ZM^{d-1}\times\NM)\otimes\CM^{M}$ is the partial isometry given by restriction of wave functions and $\widetilde{A}_\omega$ is norm limit of operators acting only on a strip $\ZM^{d-1}\times\{1,\ldots,N\}$ and being covariant in the infinite direction, that is $S_j\widetilde{A}_\omega S_j^*=\widetilde{A}_{T_j\omega}$ for $j=1,\ldots,d-1$. In the terminology of \cite{KRS,PS}, $\widetilde{A}$ lies in the ideal $\Ee\subset T(\Aa)$ of edge operators. The three C$^*$-algebras form a short exact sequence $0\to\Ee\to T(\Aa)\to\Aa\to 0$ which is used to prove the bulk-boundary correspondence.

\vspace{.2cm}

On $\Ee$ exists a densely defined trace $\widehat{\Tt}$ defined by
$$
\widehat{\Tt}
(\widetilde{A})
\;=\;
\EE\;\sum_{n_d\geq 0}
\;\Tr\,\langle 0,n_d|\widetilde{A}_\omega|0,n_d\rangle
\;,
$$
where the $0$ on the r.h.s. is the origin in $\ZM^{d-1}$. This trace is the the trace per unit volume along the boundary combined with the usual trace in the direction perpendicular to the boundary. More informations on this trace and $\widehat{\Tt}$-traceclass operators can be found in \cite{PS}. 


\end{document}